\documentclass[final,3p,twocolumn]{elsarticle}

\usepackage{epsfig}

\usepackage{amssymb}

\journal{Nuclear Instruments and Methods in Physics Research Section A}

\begin{document}

\begin{frontmatter}



\title{TORCH: Time of Flight Identification with Cherenkov Radiation}


\author[ox]{M. J. Charles}
\author[cern]{R. Forty}
\author{on behalf of the LHCb Collaboration}

\address[ox]{Department of Physics, University of Oxford, Oxford, United Kingdom.}
\address[cern]{European Organisation for Nuclear Research (CERN), Geneva, Switzerland.}

\begin{abstract}
TORCH is a time-of-flight detector concept using Cherenkov light to provide
charged particle identification up to 10\,GeV$/c$. The concept and design
are described and performance in simulation is quantified.
\end{abstract}


\end{frontmatter}


\section{Introduction}
\label{sec:intro}

Charged particle identification is critical for a heavy flavour physics experiment.
It is required both for separating signal decays from more copious backgrounds,
such as $B^0_s\rightarrow D^-_s K^+$ from $B^0_s\rightarrow D^-_s\pi^+$,
and for tagging the initial flavour of neutral mesons which may undergo mixing.
One way to accomplish this is with a Ring Imaging Cherenkov (RICH) system,
in which mass hypotheses are tested by measuring the angle ($\theta_C$) and number
of Cherenkov photons emitted by a charged particle passing through a radiator.
For a given detector geometry, the separation power between $K$ and $\pi$ hypotheses
depends on the refractive index $n$ of the radiator and the momentum $p$ of the
particle. 
Between the pion and kaon thresholds, pions are positively identified
but kaons are identified only by the absence of photons.
In principle this can provide good separation, but
it is vulnerable to background from fake tracks (e.g. due to
mistakes in pattern recognition in the tracking system), which cannot
create Cherenkov photons and so would be identified as kaons. Likewise,
protons would not be distinguished from kaons. Therefore, it is desirable
to work above the kaon threshold---and in an environment with high
occupancy in the tracking system or a large number of protons, it is
essential. 

At LHCb~\cite{bib:DetPaper}, particle identification is needed for the momentum
range from 2\,GeV$/c$ to 100\,GeV$/c$. This is currently accomplished by a
RICH system~\cite{bib:TDR:RICH} consisting of 
two detectors with three radiators. 
Two of these are gaseous: C$_4$F$_{10}$ and CF$_4$, with
refractive indices of 1.0014 and 1.0005, respectively, for visible light at STP.
These correspond to kaon Cherenkov thresholds of 9.3 and 15.6\,GeV$/c$, and
together provide better than $3\sigma$ kaon-pion separation up to 100\,GeV$/c$.
Positive kaon identification from 2--10\,GeV$/c$ requires a third radiator
with a kaon Cherenkov threshold around 2\,GeV$/c$, corresponding to
$n \approx 1.03$. This excludes conventional materials: the phase transition
from gas to liquid or solid is associated with a large jump in $(n-1)$
from $\mathcal{O}(10^{-3})$ or smaller to $\mathcal{O}(1)$. 
The low density of aerogel gives it a suitable refractive index:
LHCb's third radiator is aerogel with $n=1.029$ in air and $n=1.037$
in C$_4$F$_{10}$. However, Rayleigh scattering limits the useful light
yield: approximately 7 detected photoelectrons are expected at LHCb
for a saturated track in aerogel. In a high-background scenario, such
as the proposed LHCb upgrade
from $2 \times \, 10^{32} \, \mathrm{cm}^{-2} \, \mathrm{s}^{-1}$
to   $2 \times \, 10^{33} \, \mathrm{cm}^{-2} \, \mathrm{s}^{-1}$,
kaon-pion discrimination from aerogel would be severely
compromised~\cite{bib:young-min}.

Another method of charged particle identification is by time of
flight measurement. The principle is simple: if a particle of
mass $m$ is detected at position $x$ and time $t$ after leaving
the origin, then
\begin{equation}
  t = \frac{x}{c} \sqrt{1 + \left(\frac{m}{p}\right)^2}
  \approx \frac{x}{c} \left[ 1 + \frac{1}{2}\left( \frac{m}{p} \right)^2 \right]
\end{equation}
and so the difference in expected time between kaons and pions would be
\begin{equation}
  t_K - t_\pi \approx \frac{x}{c} \frac{1}{2p^2} \left[ m_K^2 - m_{\pi}^2 \right]
  .
\end{equation}
Therefore, to obtain $3\sigma$ separation between the two hypotheses
the time resolution $\sigma_t$ must satisfy
\begin{equation}
  \sigma_t < \frac{1}{3} \frac{x}{c} \frac{1}{2p^2} \left[ m_K^2 - m_{\pi}^2 \right]
  .
\end{equation}
If this approach were to be used for kaon identification in the range
2--10\,GeV$/c$ and the path length were $x=10$\,m, then a per-track time
resolution $\sigma_t < 12.5$\,ps would be required. A very fast
response would be needed---Cherenkov radiation would be suitable for this
purpose. If the expected light yield were 50 detected photoelectrons,
a per-photon time resolution under 90\,ps would be required.

A third method is to measure the time of propagation $\tau_{\gamma}$ over a path
length $d_{\gamma}$ of Cherenkov photons emitted by a charged particle.
In a dispersive medium, photons propagate at the group velocity $c/n_g$,
and so
\begin{equation}
  \label{eq:ng}
  n_g = c \, \frac{\tau_{\gamma}}{d_{\gamma}}
  .
\end{equation}
The relationship between $n$ and $n_g$ is a non-linear function
that depends on the medium but obeys:
\begin{equation}
  n_g = n - \lambda \frac{ \mathrm{d}n }{ \mathrm{d}\lambda}
  .
\end{equation}
After computing $n$ from $n_g$, the mass of the particle can be extracted:
\begin{eqnarray}
  \beta & = & \frac{1}{n \cos \theta_C} \\
  \Rightarrow m & = & p \sqrt{ n^2 \cos^2 \theta_c - 1 }
  .
\end{eqnarray}

The methods outlined above may be combined in an elegant way.
Consider a large, thin plane of an optically dense medium such as
quartz. Charged particles passing through the plane emit Cherenkov
photons, which propagate to the edges via total internal reflection
as in the BABAR DIRC~\cite{bib:babarNIM}.
The arrival time of a photon at the edge of the plane is the
sum of the time of flight of the track to the emission point, $t$,
and the time of propagation of the photon in the medium, $\tau_{\gamma}$. Then
\begin{equation}
  \label{eq:timeSum}
  t + \tau_{\gamma}
  = \frac{x}{c} \sqrt{1 + \left(\frac{m}{p}\right)^2} 
  + \frac{d_{\gamma} n_g}{c}
\end{equation}
where $n_g$ may be determined from
\begin{equation}
  \label{eq:timeSum_n}
  n = \frac{1}{\beta \cos \theta_C} = \frac{\sqrt{m^2 + p^2}}{p \cos \theta_C}
  .
\end{equation}
Both $\mathrm{d}t/\mathrm{d}m$ and $\mathrm{d}\tau_{\gamma}/\mathrm{d}m$
have the same sign so the two effects combine constructively.
For a given track, the first term of Eq.~\ref{eq:timeSum} is fixed but the second
depends on the direction and path length of the photon,
which must be reconstructed for each photon individually---for the
planar geometry outlined above, this can be done by measuring two
independent angles for each photon.
This is the principle behind the design of the TORCH,
and of the Belle upgrade TOP~\cite{bib:belleTop99}
and PANDA endcap DIRC~\cite{bib:pandaDirc} which inspired it.

\section{TORCH design}
\label{sec:design}

\begin{figure}[htb]
  \begin{center}
    \epsfig{file=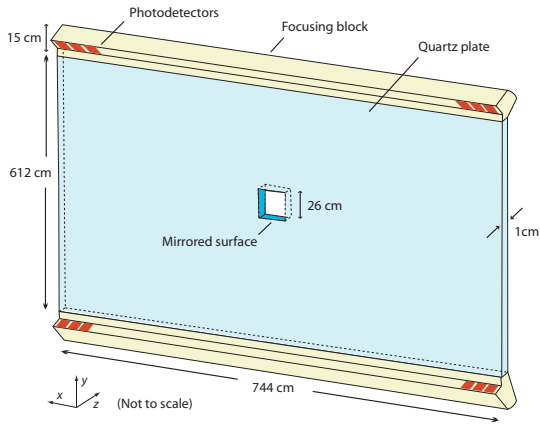,width=0.9\columnwidth}
  \end{center}
  \vspace*{-2mm}
  \caption{
    Schematic layout of the TORCH detector.
    For clarity, focusing elements have only been shown on the upper
    and lower edges.
    Photodetectors extend along the full length of each edge.
  }
\label{fig:layout}
\end{figure}

\begin{figure}[htb]
  \begin{center}
    \epsfig{file=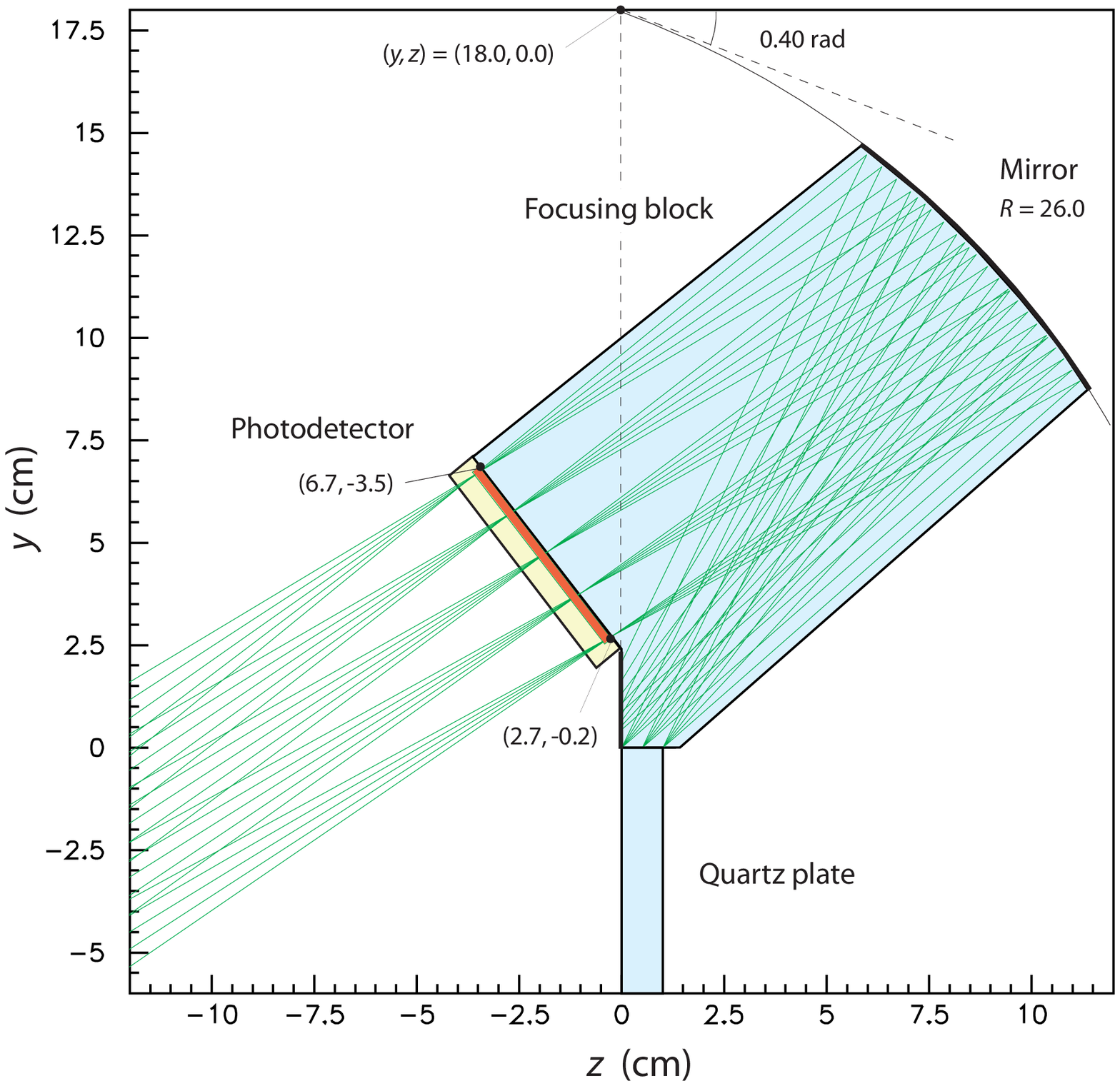,width=0.57\columnwidth}~
    \includegraphics*[width=0.41\columnwidth,viewport=325 0 505 240,clip]{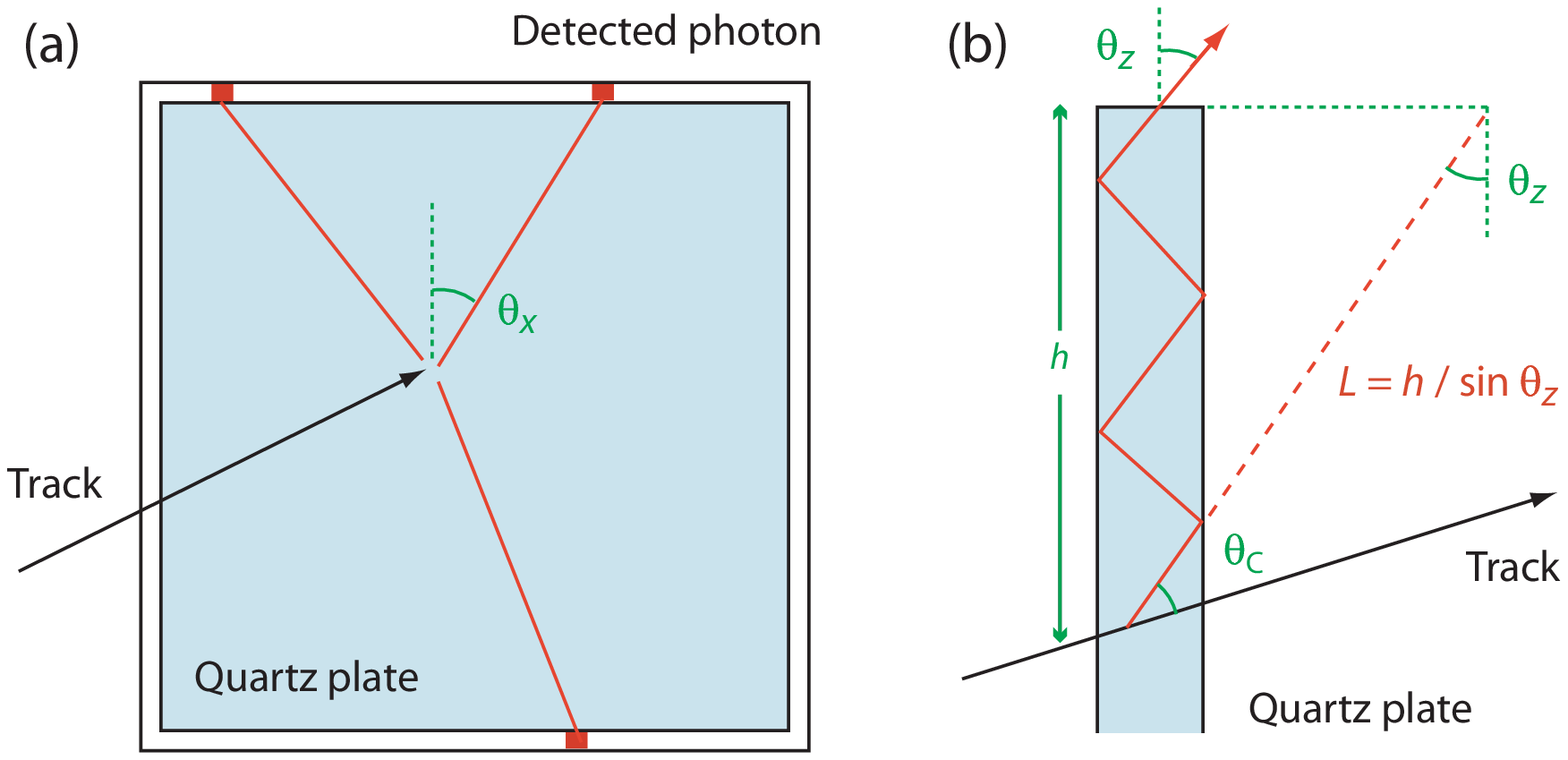}
  \end{center}
  \vspace*{-2mm}
  \caption{
    Cross section of the focusing element (left), and definition
    of the angle $\theta_z$ (right). The focusing of photons is
    shown for five illustrative angles between 450 and 850\,mrad,
    emerging at different points across the edge of the plate.
  }
\label{fig:focus}
\end{figure}

The TORCH detector consists of a
plane of quartz covering the full LHCb angular acceptance and placed
12\,m downstream of the interaction point, plus focusing elements
and photodetectors along each edge. The quartz plate has dimensions
7440\,mm\,$\times$\,6120\,mm, with a hole of 26\,mm\,$\times$\,26\,mm
at the centre for the beampipe, as illustrated in Fig.~\ref{fig:layout}.
The focusing element uses a cylindrical mirror to focus the light 
such that the distance along the photodetector plane is approximately
proportional to the angle $\theta_z$, as illustrated in
Fig.~\ref{fig:focus}. This angle is key: the overall resolution
is dominated by the pixelization uncertainty on $\theta_z$.
From simulation studies, we find that a 128-channel segmentation and
a dynamic range of 400\,mrad corresponds to an uncertainty of
0.96\,mrad on $\theta_z$ and per-photon uncertainty on the time of
propagation of $\mathcal{O}(70\,\mathrm{ps})$. The granularity
requirement in the other dimension is much looser, since the azimuthal
angle is typically measured with a lever arm of a few metres:
$\mathcal{O}(1\,\mathrm{cm})$ is sufficient match the 1\,mrad resolution.

The key property required of the photodetector is high speed: its
intrinsic time spread should be significantly better than 90\,ps.
The photon detection technology which currently gives the best
time resolution is the use of micro-channel plate photomultipliers
(MCP-PMT). They have been used in R\&D for a related concept under
development by the Belle collaboration for their upgrade~\cite{bib:BelleTOP},
with single-photon time resolutions of 35\,ps achieved.

An example of a commercially-available detector of this type
which comes close to satisfying the requirements of TORCH is the
Photonis$^{\rm TM}$ XP85022 MCP-PMT. This is available in a
1024-channel version, with an array of $32\times 32$ pixels in
a $53\times 53\,$mm$^2$ active area, and overall dimension of 59\,mm.
For the sake of developing a concrete design for TORCH we have
assumed the use of this photodetector, laid side-by-side with a pitch of
60\,mm along the edges of the plane. The anode structure of an MCP-PMT
consists simply of conductive pads to read out the charge,
and the pixellization can therefore in principle be adjusted.
For the simulation of TORCH we have assumed an array of $8\times 128$ pixels
of dimension $6.6 \, \mathrm{mm} \times 0.4 \, \mathrm{mm}$.

\section{Reconstruction and pattern recognition}

Following the approach outlined in Sec.~\ref{sec:intro} for the
design from Sec.~\ref{sec:design}, the reconstruction is
relatively straightforward: we measure the arrival time and position
of a photon on the photodetector plane, infer $\theta_z$, and calculate
its trajectory assuming it was emitted by the track at the midpoint in $z$
of its path through the quartz block. Given measurements of the
track path and momentum from a tracking system and knowledge of the optical
properties of quartz, this is sufficient information to compute all quantities
in Eq.~\ref{eq:timeSum} and~\ref{eq:timeSum_n} and extract the mass
from $t + \tau_{\gamma}$. There are two major practical challenges, though:
first, association of photons to tracks (pattern recognition); and
second, determining the time when the track left the origin.

Pattern recognition is critical: in the environment of the upgraded LHCb
luminosity, there may be $\mathcal{O}(100)$ fully reconstructed tracks
plus a large number of secondaries passing through the TORCH in an event.
However, most photon-track pairs can be rejected as unphysical. We take
advantage of the limited range of photon wavelengths to which the
photodetector is sensitive and reject candidates outside this range.
This is mathematically equivalent to a restriction on $\theta_C$:
for our geometry, this means that relevant photons lie roughly along arcs
on the photodetector plane.
Additionally, we can ignore photon-track pairs with unphysical
timing. Instead of attempting to measure the mass of the particle directly,
we assume $e, \mu, \pi, K, p$ hypotheses and test for consistency
with each in turn. Background photons whose timing is not consistent with
any of these masses can be ignored.

In Sec.~\ref{sec:intro} it was assumed that the track left the origin
at $t=0$---or equivalently, that the relative timing between the
photodetector and the initial collision is known. However, we cannot
rely on an external clock for 10\,ps-level timing---and even if we could,
the luminous region at LHCb extends a few~cm in $z$, which would add a
smearing of order a few tens of ps. We must therefore measure the track
start time ourselves. One approach would be to install a second TORCH-like
detector close to the interaction point, at the price of additional cost,
material, and time resolution. A more efficient solution is to take
advantage of the high pion multiplicity at a hadron
collider and use other tracks from the same event primary vertex to fix the
relative timing. The same reconstruction procedure described above is
used except that the last step is inverted: for each track from the
event primary vertex, the pion mass is assumed as the input and the time
elapsed $(t + \tau_{\gamma})$ is calculated from Eq.~\ref{eq:timeSum}. Subtracting
this from the measured photon detection time gives a measurement of the
track start time. For a primary vertex with $N_{\pi}$ fully reconstructed pion
tracks, the start time resolution is smaller than the per-track time
resolution by roughly $\surd N_{\pi}$.

\section{Performance}

To test the particle identification performance, we take events from the
full LHCb Monte Carlo simulation, determine the set of charged particles that
would pass through the TORCH,
simulate the emission, propagation, and detection of Cherenkov radiation in
the TORCH for these particles with a stand-alone program, then apply the
reconstruction and pattern-recognition procedure outlined above to the output.
No truth information is used for the reconstruction and pattern-recognition.
However, a number of simplifying assumptions were made: 
  perfect measurements of track position and momentum, 
  perfect extrapolation of tracks through the magnetic field of the detector,
  negligible electronics noise and cross-talk; 
  negligible spill-over between events; and
  no multiple scattering, delta ray emission, or inelastic collisions in the TORCH.
Background from certain sources such as backscatter from the calorimeters was not
available to the simulation and was also neglected.

\begin{figure}[htb]
  \begin{center}
    \includegraphics*[width=0.75\columnwidth,viewport=5 2 521 356,clip]{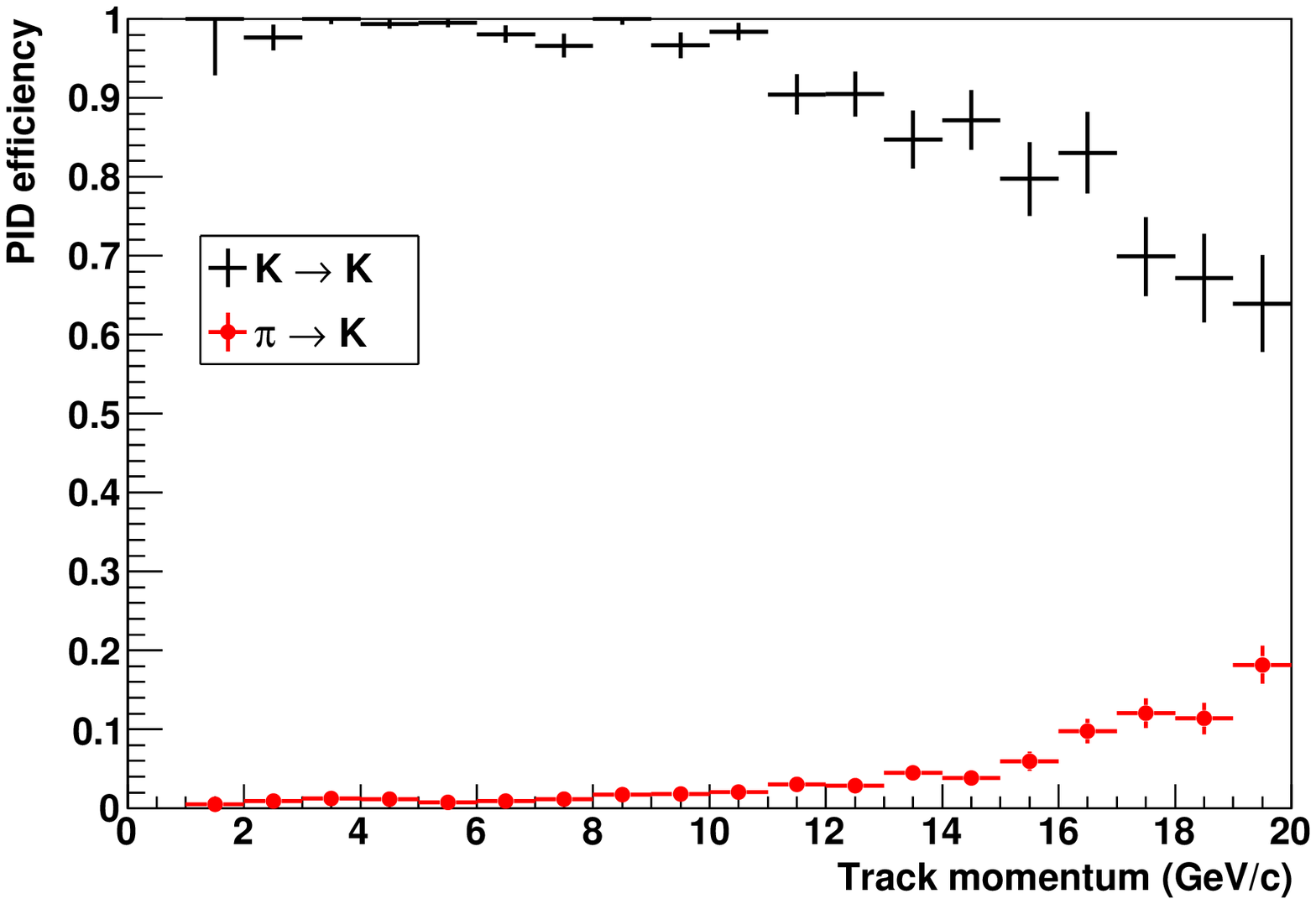}
    \includegraphics*[width=0.75\columnwidth,viewport=5 2 521 356,clip]{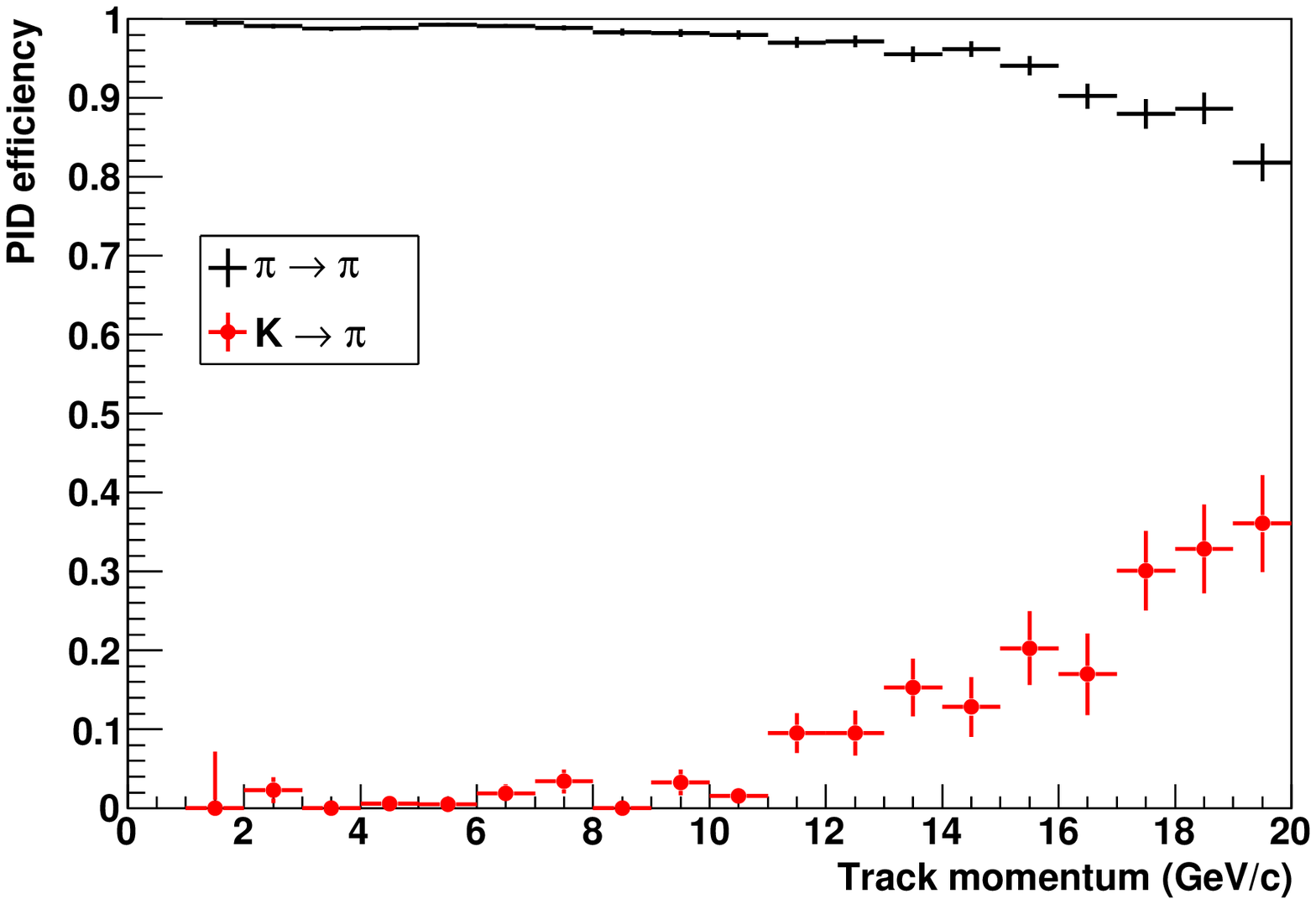}
  \end{center}
  \caption{
    Identification efficiency of the subset of well-measured charged tracks
    which are well-matched to a primary vertex.
    The plots show the efficiency for a kaon (upper) or pion (lower) track to be
    identified correctly (black) or incorrectly (red/dot), considering only the
    kaon and pion hypotheses.
  }
  \label{fig:effic-reqPV}
\end{figure}

We characterize the performance in terms of particle identification efficiencies
for those kaon and pion tracks which would be useful for physics analysis.
Specifically, we require tracks to be well-measured (with track segments found
both upstream and downstream of the magnet), well-matched to an event primary
vertex, and associated to a kaon or pion in the Monte Carlo truth information.
Tracks are identified either as kaons or as pions according to which of the
two hypotheses is found to be most consistent by the pattern recognition.
The efficiency is shown as a function of track momentum in
Fig.~\ref{fig:effic-reqPV} for events simulated at a luminosity of
$L = 2 \times \, 10^{32} \, \mathrm{cm}^{-2} \, \mathrm{s}^{-1}$.
The efficiency for correct identification is $> 95\%$ up to
10\,GeV$/c$, dropping at higher momentum as the time difference between
kaon and pion hypotheses diminishes.

\section{Conclusions}

The TORCH concept is intended to provide charged particle identification in
the momentum range 2--10\,GeV$/c$ in a high-rate environment.
It has been shown that the pattern-recognition problem is solvable 
and that target performance can be reached under simplified conditions.
The challenge is now to test whether this performance can be maintained
under realistic conditions and at
$L = 2 \times \, 10^{33} \, \mathrm{cm}^{-2} \, \mathrm{s}^{-1}$.
This will require prototype tests and more detailed simulation.
A revised, modular design with a smaller quartz plane is also
under investigation.

\section{Acknowledgements}

The TORCH concept has evolved from ideas that have been demonstrated in the 
BABAR DIRC, and related developments that are under study by the PANDA and Belle
collaborations.  It is a pleasure to thank Bjoern Seitz for fruitful discussion,
and our colleagues in the LHCb RICH group for their interest.




%



\end{document}